\documentclass[preprint]{aastex63}

\def\be{\begin{equation}}
\def\ee{\end{equation}}
\def\ba{\begin{eqnarray}}
\def\ea{\end{eqnarray}}
\newcommand{\msun}{\ifmmode\mbox{M}_{\odot}\else$\mbox{M}_{\odot}$\fi}
\newcommand{\rsun}{\ifmmode\mbox{R}_{\odot}\else$\mbox{R}_{\odot}$\fi}
\newcommand{\degrees}{\ifmmode^{\circ}\else$^{\circ}$\fi}
\newcommand{\degree}{\ifmmode^{\circ}\else$^{\circ}$\fi}
\newcommand{\amin}{\ifmmode^{\prime}\else$^{\prime}$\fi}
\newcommand{\asec}{\ifmmode^{\prime\prime}\else$^{\prime\prime}$\fi}

\shorttitle{AO327 Pulse Profile Catalog}
\shortauthors{Deneva et al.}

\begin{document}

\title{The AO327 Drift Survey Catalog and Data Release of Pulsar Detections}

\author[0000-0003-1226-0793]{J.~S.~Deneva} 
\affiliation{George Mason University, 4400 University Dr, Fairfax VA 22030, USA}

\author{M. McLaughlin}
\author{T. E. E. Olszanski}
\author{E. F. Lewis}
\author{D. Pang}
\affiliation{West Virginia University, P.O. Box 6315, Morgantown, WV 26506, USA}

\author{P. C. C. Freire}
\affiliation{Max-Planck-Institut f$\ddot{u}$r Radioastronomie, Bonn, Germany}

\author{M. Bagchi}
\affiliation{The Institute of Mathematical Sciences, Chennai, India 600113}
\affiliation{Homi Bhabha National Institute, Mumbai, India 400094}

\author{K. Stovall}
\affiliation{University of New Mexico, Albuquerque, NM 87131, USA}


\begin{abstract}
The AO327 drift survey for radio pulsars and transients used the Arecibo telescope from 2010 until its collapse in 2020. AO327 collected $\sim 3100$ hours of data at 327~MHz with a time resolution of 82~$\mu$s and frequency resolution of 24~kHz. While the main motivation for such surveys is the discovery of new pulsars and new, even unforeseen, types of radio transients, they also serendipitously collect a wealth of data on known pulsars. We present an electronic catalog of data and data products on 206 pulsars whose periodic emission was detected by AO327 and are listed in the ATNF catalog of all published pulsars. The AO327 data products include dedispersed time series at full time resolution, average (``folded") pulse profiles, Gaussian pulse profile templates, and an absolute phase reference that allows phase-aligning the AO327 pulse profiles in a physically meaningful manner with profiles from data taken with other instruments. We also provide machine-readable tables with uncalibrated flux measurements at 327~MHz and pulse widths at 50\% and 10\% of the pulse peak determined from the fitted Gaussian profile templates. The AO327 catalog data set can be used in applications like population analysis of radio pulsars, pulse profile evolution studies in time and frequency, cone and core emission of the pulsar beam, scintillation, pulse intensity distributions, and others. It also constitutes a ready-made resource for teaching signal processing and pulsar astronomy techniques.  
\end{abstract}

\section{Introduction}

The AO327 drift survey for pulsars and transients began in 2010 when a long repair of the Arecibo telescope restricted its movement. After the repair was completed, the survey continued for ten years as a filler project during subsequent down times, weather-related shut-downs of normal tracked observations, and otherwise unallocated telescope time. The goal of AO327 was to cover the entire Arecibo sky (declinations, $\delta$, from $-1\degree$ to $38\degree)$ with the exception of the region within $5\degree$ of the Galactic plane where a higher observing frequency is optimal and was observed by the concurrent {Pulsar-ALFA (PALFA) survey using the seven-beam Arecibo L-band Feed Array (ALFA) at 1.4~GHz} (e.g. \citealt{Cordes06} , \citealt{Deneva09}, \citealt{Lyne17}, \citealt{Patel18}, \citealt{Parent22}). \cite{Deneva13} presented an initial batch of 24 pulsars newly discovered by the AO327 survey, including three millisecond pulsars (MSPs) and five binary systems. Subsequent investigations included the discovery and timing of two eccentric double neutron star systems (\citealt{Martinez15}, \citealt{Martinez17}) and an eccentric neutron star - white dwarf system with an unusual evolution scenario (\citealt{Antoniadis16}, \citealt{Stovall19}). In addition, \cite{Martinez19} reported the discovery of six recycled pulsars and \cite{Deneva16} developed an algorithm for automatic identification of astrophysical single pulses that facilitated the discovery of 22 pulsars and rotating radio transients (RRATs). Three MSPs discovered by AO327 are exceptionally stable rotators and are included in the NANOGrav pulsar timing array project whose goal is to detect nanohertz gravitational waves (\citealt{Nanograv15yrA}, \citealt{Nanograv15yrB}). 

Our motivation for releasing a catalog of AO327 known pulsar detections is three-fold. AO327 was the largest Arecibo pulsar survey both in sky area covered and in total observing time. Its data set, like the data sets of all surveys conducted with Arecibo, is part of the legacy of the Arecibo Observatory whose science operations will cease in 2023. As the price of computer storage space has decreased and data throughput over the internet has increased significantly over the years, it is now feasible to make publicly available the most interesting portions of survey data: those containing known pulsars and transients, and various data products that would be of interest to the scientific community. We would like to inspire other surveys to do the same and to begin a discussion about establishing a central repository of data and data products on known pulsars that can be searched as easily as the ATNF catalog.

The work on the AO327 pulse profile catalog is also an essential precursor for a future population analysis paper. While the current work focuses on  pulsars detected via their periodic emission, a population study will also include RRATs that were detected only via a single-pulse search. Another motivation for releasing this catalog is that there is a wide variety of research and teaching projects our data could be used for, including some outside of the immediate interests or priorities of the AO327 team. These include for example studying the pulse intensity distributions of pulsars, cone and core emission and beam components, teaching a signal processing course using pulsar data, and many others. The AO327 survey accumulated a large backlog of data that is still being analyzed and searched for new pulsars. In the meantime, the scientific community can already take advantage of the data we have accumulated on known pulsars. 

In Section~\ref{sec:obs} we describe survey observations and sky coverage. In Section~\ref{sec:prep} we focus on locating and extracting the short data chunks covering the positions of known pulsars from the 1-hour survey raw data files. In Section~\ref{sec:analysis} we describe our pipeline for processing the data chunks: radio frequency masking, dedispersion, and obtaining an averaged pulse profile for each pulsar. Section~\ref{sec:catalog} lists all data products and results provided as part of the catalog and the further analysis steps for producing them. Section~\ref{sec:non} discusses the reasons why some of the known pulsars in our survey coverage area were not detected. We conclude in Section~\ref{sec:conclusion}.

\section{Observations}\label{sec:obs}

The AO327 drift survey for pulsars and transients operated from 2010 to 2020 and was cut short when the Arecibo telescope collapsed and was irretrievably lost. Survey observations used otherwise unallocated time, or were carried out during telescope repairs and weather-related shutdowns when tracked observations were not feasible due to constraints or stoppage of telescope movement. During drift observations the telescope was typically parked at an azimuth of either 0\degrees\ or 180\degrees, with elevation increasing by 
15\arcmin\ between strips of sky adjacent in $\delta$. This spacing corresponds to the Arecibo {full-width half-maximum (FWHM)} beam diameter at 327~MHz. A small number of observations were carried out at other azimuth values during hurricane-related shutdowns when the telescope was in its stow position with the azimuth arm oriented in the direction of prevailing wind. 

\begin{figure}
\begin{center}
  \includegraphics[width=0.90\textwidth]{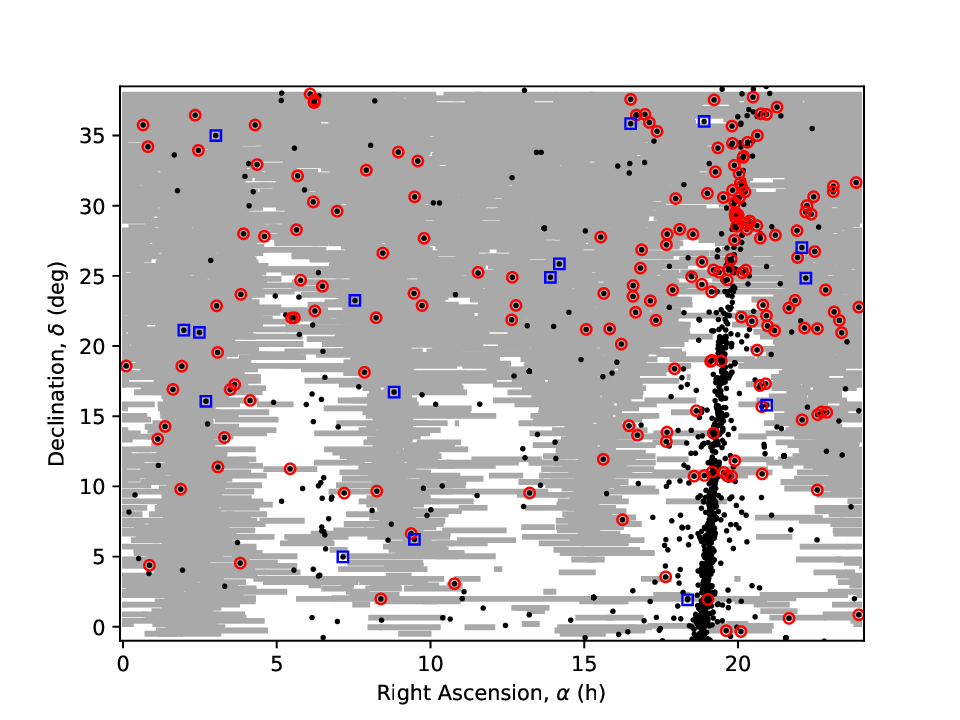}
\caption{Sky coverage of the AO327 survey with the PUPPI spectrometer is shown in grey. Known pulsars from the ATNF catalog are shown as black points. {Pulsars included in the AO327 catalog are indicated with blue squares if they were discovered in AO327 PUPPI data by their periodic emisssion and with red circles if they are otherwise known pulsars listed in the ATNF catalog. RRATs and slow pulsars discovered by AO327 via a single-pulse search are not included in the present work if no periodic emission was detected in AO327 survey data.} The concentration of pulsars near right ascension of $\sim 18-20$~h corresponds to the inner Galactic plane.\label{fig_coverage}}
\end{center} 
\end{figure}

From 10 February 2010 to 16 March 2014 we used the Mock spectrometer\footnote{\tt http://www.naic.edu/\~{}phil/hardware/pdev/pdev.html}, and from 17 March 2014 to 8 August 2020 we used the newer PUPPI backend (Puerto Rico Ultimate Pulsar Processing Instrument), which was a clone of the similarly named and now decommissioned GUPPI at Green Bank\footnote{\tt http://safe.nrao.edu/wiki/bin/view/CICADA/GUPPISupportGuide}. PUPPI provided better sensitivity compared to the older Mock backend. \cite{Deneva13} present a detailed comparison of the sensitivities of AO327-Mock and AO327-PUPPI with other pulsar surveys. The bulk of AO327 data were taken with PUPPI, and it was our goal to survey the entire Arecibo sky {($-1\degrees < \delta < 38\degrees$) outside of the Galactic plane ($| b | > 5\degrees$)} using this backend, including the portions previously covered with the Mock spectrometer. Although we were unable to complete this goal in its entirety because of the collapse of the telescope, we managed to cover most of the Arecibo sky with PUPPI. The data products and measurements in the AO327 catalog are from the PUPPI data set, and Figure~\ref{fig_coverage} shows the extent of its sky coverage. 

Drift data were taken as contiguous observations of one hour each to allow periodic {retuning of the intermediate frequency and local oscillator (IF/LO) stages of the signal path and rebalancing backend power levels} as ambient and radio frequency interference (RFI) conditions changed. Data were recorded {in a summed-polarizations mode} and in an 8-bit PSRFITS format \citep{PSRFITS}. After an observation was completed, the data were converted to 4-bit PSRFITS. This conversion is also employed by the PALFA survey; it preserves the data dynamic range and sensitivity while decreasing storage requirements by half. We used a PUPPI mode with a bandwidth of 100~MHz, 4096 channels, and a sampling time of 81.92~$\mu$s. However, we only recorded data from the 2816 channels covering the usable 68.75~MHz bandwidth of the 327~MHz receiver. A total of 3091 hours of PUPPI data were recorded, and $\sim 62\%$ of the data have been completely processed and searched for new pulsars and transients. 

\section{Data Preparation}\label{sec:prep}

The full PUPPI data set consists of thousands of PSRFITS data files covering strips of sky, some of which partially overlap. Our first task is to {find and extract the portions of data files covering the positions of known pulsars.} PSRFITS files contain a table of the position of the telescope beam center that is recorded at regular intervals throughout an observation, in our case every 8192 samples, corresponding to 0.67~s. We extracted right ascension ($\alpha$) and declination ($\delta$) coordinates recorded every $\sim 10$~s from these tables for all data files. Next we used these positions along with the Arecibo beam diameter at 327~MHz ($\sim 15\amin$) to construct contours corresponding to the strips of sky traced out by the FWHM telescope beam. We then created a MySQL database of these geometric areas and PSRFITS header information for all data files. We downloaded a copy of the {ATNF pulsar catalog\footnote{\tt https://www.atnf.csiro.au/research/pulsar/psrcat} (v1.70; \citealt{ATNF})} containing all known pulsar positions and used that along with the MySQL {\tt MBRContains} function in a query yielding a list of known pulsars and  corresponding raw data files covering their positions. We located 351 known pulsars' positions within the sky coverage of AO327, some of which were observed more than once in raw data files covering overlapping or adjacent strips of sky. 

The FWHM beam diameter of Arecibo at 327~MHz is $D_{\rm beam} = 15\amin$, and the sidereal rate is $\Delta r = 15\amin$~min$^{-1}$. A source at $\delta$ passing through the beam center has a transit time of 
\be
t_{\rm tr} \approx \frac{D_{\rm beam}}{\Delta r~{\rm cos}~\delta}.
\ee
A source on the celestial equator would transit the beam in 60 seconds, while a source at 38\degrees, the highest $\delta$ accessible to Arecibo, would transit in 76~s. Since our drift observations are untargeted, most pulsars do not transit through the beam center but at some offset in $\delta$, such that their transit time through the FWHM beam is $< t_{\rm tr}$. However, for many pulsars emission can still be detected when they are outside the FWHM as well. 

When blindly searching the data for new pulsars we split raw data files into consecutive 60~s chunks overlapped by 30~s \citep{Deneva13}, and process each data chunk as a separate ``pointing". For the purpose of extracting data chunks around all known pulsars covered by survey observations, we use a data chunk length of 64~s, corresponding to the transit time through the beam center for a source at $\delta \approx 19\degrees$, the midpoint in $\delta$ for the AO327 survey area. For each pulsar we extract a data chunk such that it covers a $\sim 15\amin$ strip of sky centered on the $\alpha$ listed in the ATNF catalog for the known pulsar. If a pulsar was within the telescope beam at the start or end of a 1-hour raw data file, the extracted data chunk is asymmetric with respect to the pulsar's $\alpha$ and its duration is $< 64$~s. 

\section{Data Analysis}\label{sec:analysis}

We analyzed each extracted data chunk with tools from PRESTO \citep{Ransom02}. 

\subsection{Radio Frequency Interference Masking}\label{sec:rfi}

First we masked radio frequency interference (RFI) using {\tt rfifind}. Statistics were computed on consecutive blocks of samples 2~seconds long and preserving the full resolution in frequency. The smallest unit of RFI masking was therefore a 2-second stretch of a single frequency channel. The largest was either a range of frequency channels for the entire duration of the data chunk, or a continuous time range for the entire range of frequency channels. The portions of the time-frequency plane to be masked were determined by their deviation from the median values of the variance, mean power, or total power for the data chunk. {Two additional criteria were: 1) a frequency channel was completely masked if at least 30\% of the 2-second blocks within it were already flagged for masking; 2) a 2-second time interval was completely masked if at least 70\% of the channels within it were already flagged for masking.} When the RFI mask was applied during data processing steps downstream, completely masked channels were ignored (set to zero), while other masked portions of data were replaced with a local average. 

\subsection{Dedispersion}

Electromagnetic radiation traveling through an ionized gas medium experiences a frequency-dependent delay according to the cold plasma dispersion relation. {The pulse arrival time delay $\Delta t_{DM}$ has a $\propto \nu^{-2}$ dependence on the electromagnetic frequency $\nu$ compared to an infinite frequency (see e.g. Chapter 4 of \citealt{Handbook})}.
After applying the RFI mask to the time-frequency plane of the data chunk, we shifted all frequency channels with respect to the highest-frequency edge of the band according to the cold plasma dispersion relation, and then summed them using PRESTO's {\tt prepdata} to produce a one-dimensional dedispersed time series that is part of our catalog files for all pulsars detected by AO327. 

\subsection{Pulse Profiles}\label{sec:profiles}

We used {\tt prepfold} to produce a dedispersed and ``folded" pulse profile of total intensity vs. pulse phase.
To minimize the chance of missing weak detections, we processed each data chunk with {\tt prepfold} in three ways. First we folded it with DM and pulse period $P$ fixed at the ATNF values and a period derivative $\dot{P}$ fixed at zero. This generally yields good detections for slow pulsars with relatively recent published DM and P. Next we redid this stage with the DM fixed, while allowing {\tt prepfold} to search narrow ranges around the ATNF value of $P$ and $\dot{P} = 0$. This works better for slow pulsars with old $P$ measurements in the ATNF catalog, and for binary millisecond pulsars (MSPs) whose apparent $P$ and $\dot{P}$ vary with time due to the Doppler shift induced by orbital motion. Finally, we repeated the process allowing a DM as well as a $P$ and $\dot{P}$ search. This lets us detect discrepancies between published values of DM and the best-fit values from our observations, but the DM best-fit is more prone to being affected by imperfectly excised RFI. Such discrepancies may result either from secular changes of the pulsar DM due to turbulence in the interstellar medium, or from a DM measurement made at a higher radio frequency being less precise than what we can obtain from AO327 data.

Some MSPs in tight binary systems experience changes in their apparent $P$ and $\dot{P}$ large enough that neither of the above methods is sufficient to detect them: the ranges {\tt prepfold} can search for these parameters are too small. In the absence of a detection at this point, we performed a Fast Fourier Transform search for periodic signals on the dedispersed time series with {\tt accelsearch}, the same tool that is routinely used when blindly searching for new pulsars. If there was a reasonable match to P in the candidate signal list produced by {\tt accelsearch}, we folded the data chunk with the parameters of that candidate. 

After the steps of RFI masking, dedispersion, and folding, we obtained 239 detections of 206 unique pulsars. The parameters of these detections and additional data products derived from them and described below comprise the AO327 catalog.

\section{Catalog}\label{sec:catalog}

Here we describe the steps leading to each of the data products and results we provide as part of the AO327 pulsar catalog\footnote{\tt http://ao327.nanograv.org}. A more detailed listing of all files and their formats can be found in Appendix~\ref{sec:catfiles}.


\subsection{Improvements of ATNF Pulsar Parameters}

Because $\Delta t_{DM} \propto \nu^{-2}$ but sampling times are hardware-limited, observations at a lower frequency can determine a pulsar's DM with better precision than observations at a higher frequency. We found that for some pulsars discovered at higher radio frequencies than 327~MHz we can provide a more precise DM value than the one currently listed in the ATNF catalog. For RRATs, or objects generally only detectable through their single pulses, a value of $P$ listed in the ATNF catalog may have low precision because it was determined from only a few single pulses. For RRATs with a higher number of pulses or for slow pulsars, a listed $P$ may be a harmonic or sub-harmonic of the actual period. Table~\ref{tab:dms} shows ATNF values of $P$ and DM for a number of known pulsars and improved measurements from AO327 data. 

\begin{deluxetable}{lcll}
\label{tab:dms}
\tablecaption{Pulsar parameters from the ATNF catalog and improved values from AO327 data. Rotational periods $P$ are in seconds and dispersion measures DM are in pc~cm$^{-3}$.}
\tablehead{\colhead{Pulsar} & \colhead{Parameter} & \colhead{ATNF} & \colhead{AO327}}
\startdata
J0241+16 & DM & \phn16(-) & \phn20.4(5) \\
J0301+35 & $P$ & \phn0.1468(-) & \phn\phn0.568037(32) \\
J0355+28 & $P$  & \phn0.0943(-) & \phn\phn0.364946(32) \\
J1354+2454 & DM  & \phn20(-) & 20.6(3) \\
J1354+2454 & $P$ & \phn\phn6.27(-) & 0.85080(19) \\
J1921+34 & DM & \phn85(5) &  \phn88.95(23) \\
J1921+34 & $P$  & \phn\phn1.444(-) & \phn\phn1.45132(97) \\
J1931+30 & DM & \phn53.0(2) & \phn53.75(19) \\
J1937-00 & DM & \phn68.6(-) & \phn68.02(4) \\
J1939+10 & DM & \phn90(8) & \phn73.99(37) \\
J1939+10 & $P$  & \phn\phn2.31(3) & \phn\phn2.31144(44) \\
J1947+10 & DM & 149(30) & 128.76(18) \\
J2002+30 & DM & 187.80(11) & 184.06(14) \\
J2023+2853g & DM & \phn22.8(-) & \phn22.742(4) \\
J2047+1053 & DM & \phn34.6(-) & \phn34.632(2) \\
J2055+15 & DM & \phn40.2(-) & \phn40.415 \\
J2116+3701 & DM & \phn43.7(2) & \phn44.03(5) \\
J2151+2315 & DM & \phn23.6(2) & \phn20.92(19) \\
J2243+1518 & DM & \phn42.13(37) & \phn39.445(96) \\
\enddata
\end{deluxetable}


\subsection{Pulse Profiles with Phase References}\label{sec:phaseref}

The procedure described in Section~\ref{sec:profiles} yields folded pulse profiles where the zero phase corresponds to the time stamp at the start of the data chunk we ``folded". That choice of zero phase is arbitrary and has no physical significance. On the other hand, pulse arrival times do have physical significance due to the clock-like regularity of pulsar pulses. Some MSPs are exceptionally stable rotators and are used in Pulsar Timing Array (PTA) projects like NANOGrav, {the International Pulsar Timing Array (IPTA), and the Parkes Pulsar Timing Array (PPTA)} for the detection of nanohertz gravitational waves that recently announced evidence of the spatial correlations that are expected from a stochastic background of gravitational waves (e.g. \citealt{Nanograv15yrA}, \citealt{Nanograv15yrB}, \citealt{EPTA}, \citealt{PPTA}). Observations of other subsets of pulsars, also hinging on recording their pulse arrival times and pulse profile shapes, are used to study relativistic effects in binary systems, stress-test the theory of relativity, and compare its performance under these extreme conditions with alternative theories of gravity (using double neutron star systems, \citealt{Weisberg16,Kramer21} and the pulsar PSR~J0337+1715, located in a triple star system, \citealt{Archibald18,Voisin20}). Other applications include (and are not limited to) measuring neutron star masses and constraining the equation of state of super-dense matter (\citealt{Demorest10,Antoniadis13,Arz18,Fonseca21}), modelling the neutron star  beam and magnetosphere (\citealt{Wahl23}, \citealt{Rankin22}), and tests of scalar-tensor theories of gravity (\citealt{Bhat08,Freire12}).

In order to enable the community to derive maximum benefit from the AO327 catalog with a minimum amount of effort, we must ``rotate" in phase our folded profiles so that the zero phase corresponds to a pulse time-of-arrival (TOA). First we used {\tt pygaussfit.py} from PRESTO to fit the pulse shape with one or more Gaussians and obtain an analytical profile template. Next we used {\tt zerophase.py} to Fourier transform the template. By convention the zero phase was set to the phase of the fundamental component in the resulting harmonic train in Fourier space. The template was then rotated by that phase (``zerophased"). We then applied this rotated template to the folded pulse profile and also rotated the latter by the appropriate phase. 

The remaining step is to calculate a Modified Julian Day (MJD) absolute phase reference for our observing frequency and observatory location. We did this by extracting one TOA from the data chunk using the Gaussian template and {\tt get\_TOAs.py} from PRESTO; this TOA is the phase reference for the zerophased pulse profile and template.

Providing a zerophased pulse profile together with a phase reference makes it easier to solve a frequent problem encountered by pulsar researchers: how to align pulse profiles from data taken at different energy bands in a physically meaningful manner. Pulsar timing solutions and folded pulse profiles are often published without a phase reference. One typically is in possession of raw data at one energy band and an ephemeris and phase reference obtained from these data, and the only way to achieve the correct alignment is to track down raw data from all energy bands of interest and fold them with the same ephemeris and phase reference. This often involves installing and learning how to use software packages specific to processing pulsar data from each energy band. 

\begin{figure}
\begin{center}

\includegraphics[width=0.75\textwidth]{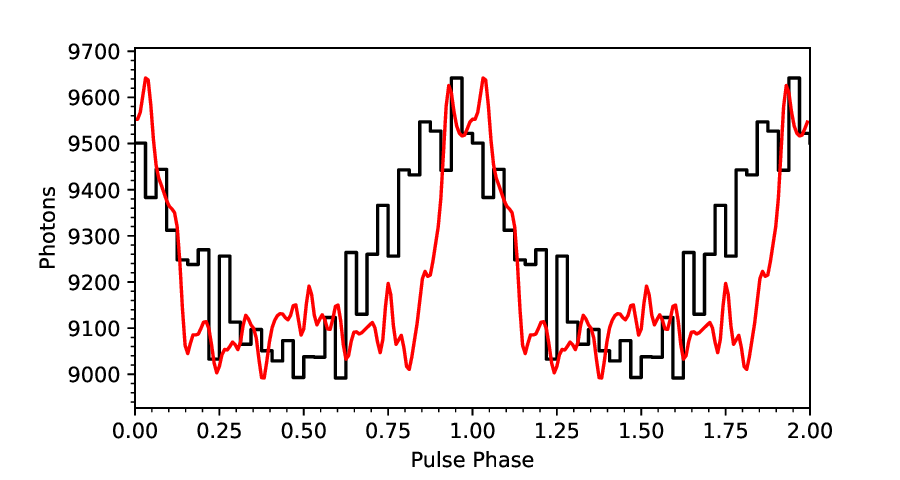}
\includegraphics[width=0.75\textwidth]{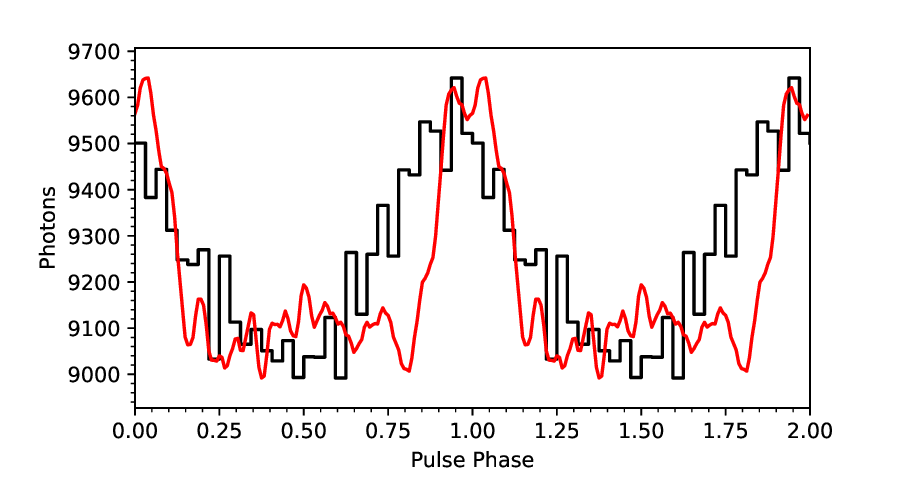}

\caption{Phase-aligned NICER (black) and Arecibo 327~MHz (red) pulse profiles of PSR~J0751+1807. The radio profiles are in arbitrary units and were normalized to the same peak height as the NICER profiles. The radio pulse profile at the top was made by folding both data sets with the pulsar ephemeris from \cite{Guillot19}. In the bottom panel the NICER data set was folded in the same way, while the AO327 data set was folded using the pulsar DM from the ATNF catalog, and the best-fit local $P$ and $\dot P$  as determined by \texttt{prepfold}. Two full rotations are shown for clarity. \label{fig_0751}}
\end{center} 
\end{figure}

Alternatively, one may take the AO327 phase reference and zerophased profile for a pulsar, and use that phase reference together with an existing ephemeris to re-fold only one's own data. Figure~\ref{fig_0751} compares profile alignments obtained with the two methods. The top panel shows NICER X-ray and AO327 radio pulse profiles of the MSP J0751+1807 aligned by folding both data sets with the same ephemeris. The bottom panel shows the NICER pulse profile folded in the same way, using the AO327 catalog phase reference, superimposed on the existing AO327 catalog zerophased pulse profile which was folded using the DM value from the ATNF catalog and the best local $P$ and $\dot{P}$ as determined by {\tt prepfold}. The radio profile's substructure is sharper when the data are folded with the ephemeris as that takes into account orbital and astrometric parameters determined over years of pulsar timing. However the profile alignment between the two energy bands is consistent, and that is often the main point of interest when studying pulsar magnetospheres and emission beam geometry.

\subsection{Pulse Widths and Duty Cycles}

Table~\ref{tab:pulsewidths} lists the pulse width and duty cycle at 50\%\ and 10\%\ of the peak for all detected pulsars, as measured from the Gaussian templates fitted to folded pulse profiles. Using a pulsar's observed FWHM pulse width with the radiometer equation when possible (see e.g. Section~\ref{sec:non}, Equation~\ref{eqn:smin}) gives more accurate results than assuming a fixed duty cycle for all pulsars. 

\subsection{Uncalibrated Flux Densities}\label{sec:flux}

AO327 data were recorded as total intensity without using a calibration diode. Providing uncalibrated flux densities in our catalog of detected known pulsars is still useful: 138 out of 206 pulsars in our catalog do not have flux densities at 300--350~MHz listed in the ATNF catalog. Of these 138, 39  have no flux density listed at any frequency $<1$~GHz, and 26 have no flux density listed at all. Having at least one measurement at a lower frequency would improve the spectral index estimate for pulsars with two or more existing measurements at higher frequencies, and would make estimating the spectral index possible for pulsars with just one other measurement. 

\begin{figure}
\begin{center}
\includegraphics[width=0.75\textwidth]{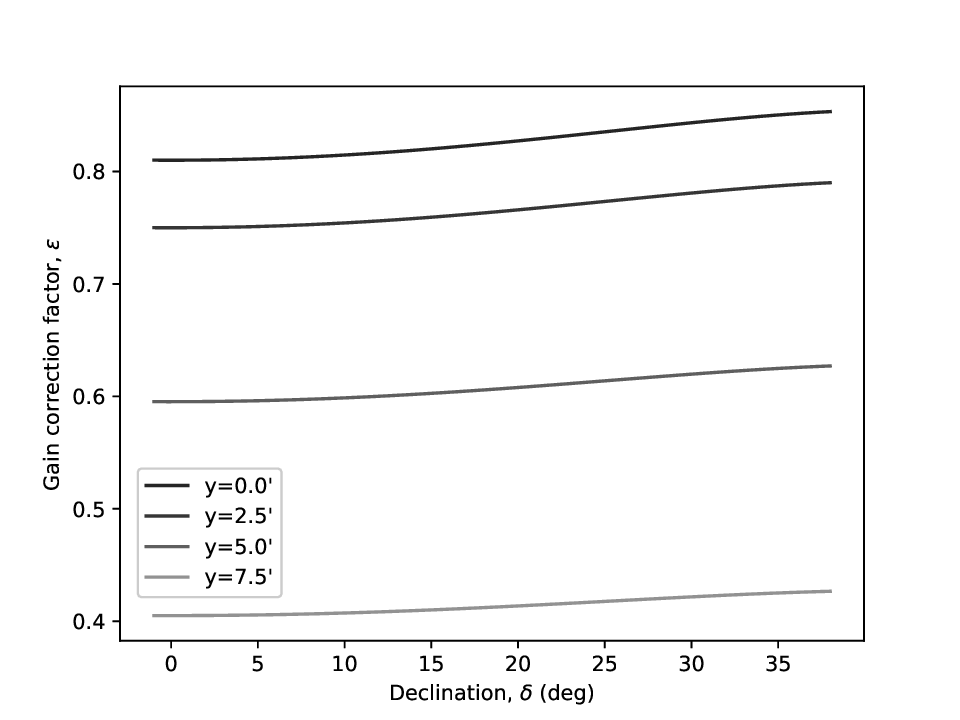}
\caption{The gain correction factor ($\epsilon$) vs. $\delta$ for several pulsar tracks at different offsets in $\delta$ ($y$) from the beam center, symmetric with respect to the beam axis aligned with the meridian, and with a length equal to the beam diameter ($15\amin$). 
\label{fig_eps}}
\end{center} 
\end{figure}

As a source transits the beam, the instantaneous gain changes as a function of the offset from the beam center. We used the same simple geometric approach as \cite{Lynch13} and modeled the telescope beam as a two-dimensional Gaussian. The nominal gain $G$ at boresight was then modified by a factor $\epsilon$ to account for the reduced and variable gain along the pulsar's transit path through the beam such that 
\be
\epsilon = \frac{1}{T_{\rm obs}} \int_0^{T_{\rm obs}} e^{-r^2\left(t\right)/f^2} dt,\label{eqn:epsilon}
\ee
where $r(t)$ is the radial offset from the beam center and $f = D_{\rm beam}/\left(2~ln~2\right)$. The normalization constant $1/T_{\rm obs}$ was chosen by requiring that $\epsilon = 1$ for a source that stays at the beam center for the entire observation time $T_{\rm obs}$. We generalized the \cite{Lynch13} expression for $r(t)$ in order to account for cases where the pulsar is within the beam at the start or end of a 1-hour raw data file. Therefore, $r^2\left(t\right) = y^2 + \left( x_0 - \dot{x}~t\right)^2$, where $x_0$ is the starting angular offset from the beam center along an axis parallel to the celestial equator, $y$ is the constant angular offset from the beam center along an axis aligned with the meridian, and $\dot{x} = \Delta r~cos~\delta$ is the angular rate of transit through the beam. Figure~\ref{fig_eps} illustrates the dependence of $\epsilon$ on $y$ and $\delta$ for the range of declinations covered by the AO327 survey. For a source on the celestial equator traversing the FWHM beam through its center, $\epsilon \approx 0.81$.

According to the radiometer equation {the mean, period-averaged} flux density of a pulsar then becomes 
\be
S = \frac{\beta~\left(S/N\right)_{\rm prof}/n_{\rm bin}~T_{\rm sys}}{\epsilon~G~\sqrt{N_p~\Delta\nu~T_{\rm obs}/n_{\rm bin}}},\label{eqn:s}
\ee
where $\beta = 1$ is the digitization factor for PUPPI; $(S/N)_{\rm prof}$ is the pulse profile signal-to-noise integrated over all bins; $G$ is the nominal gain at boresight; $N_{\rm p} = 2$ is the number of polarizations summed; $\Delta\nu$ is the bandwidth; $n_{\rm bin}$ is the number of bins in the folded pulse profile; and $T_{\rm sys} = T_{\rm rec} + T_{\rm sky}$ is the system temperature consisting of a receiver and a sky term. 

{For the 327~MHz receiver $T_{\rm rec} = 113$~K was well-controlled, measured daily, and reported to within a degree. We therefore assumed an uncertainty of 1~K. The telescope gain was $10\pm1$~K/Jy before hurricane Maria battered Puerto Rico on 16 Sep 2017 (MJD 58012) and $8\pm1$~K/Jy after. The step chainge in gain was caused by a couple of large pieces of debris falling into the telescope dish due to the storm. The impact caused warping in the dish surface and the curvature was not readjusted between the storm and the telescope collapse in 2020.

Sky temperature generally follows a power-law dependence on observing frequency such that $T \propto \nu^{\alpha_{\rm t}}$, with a typical spectral index $\alpha_{\rm t} \sim -2.7$ \citep{Gervasi08}. In order to estimate $T_{\rm sky}$ at 327~MHz for each pulsar's position we use the 408~MHz sky temperature map of \cite{Haslam82} in its digitized version \citep{Haslam_digitized} and the spectral index map of \cite{EDA2}. The latter was in turn produced by applying the power-law dependence to sky temperature measurements at 159~MHz and the aforementioned digitized 408~MHz map. \cite{EDA2} report an uncertainty of $< 0.5$~K for the map of sky temperature at 159~MHz. \cite{Haslam_digitized} report a zero-level uncertainty of 3~K and a calibration uncertainty of 5\% for the map of sky temperature at 408~MHz. We used these values when performing error propagation for $T_{\rm sky}$ at 327~MHz. Typical $T_{\rm sky}$ values are $100-150$~K for pulsars in or near the Galactic plane and $40-60$~K for off-plane pulsars detected in the AO327 survey.}

We calculated $(S/N)_{\rm prof}$ using {\tt sum\_profiles.py} from PRESTO with the Gaussian template and folded pulse profile. This tool locates on-pulse and off-pulse bins in the profile, rescales the profile so that it has an off-pulse mean of zero and off-pulse standard deviation of unity, and integrates the profile. {Since we are integrating over $n_{\rm bin}$ profile bins, the uncertainty of $(S/N)_{\rm prof}$ is $\sqrt{n_{\rm bin}}$.}

{We also took into account the uncertainty in the pulsar position because the gain modification factor $\epsilon$ depends on the position offset between the beam center and the pulsar, which varies during the observation. If a position uncertainty was not available in the ATNF catalog we assumed an uncertainty of a degree if a coordinate was listed to the nearest degree and an arcminute if it was listed to the nearest arcminute.} 

\begin{figure}
\begin{center}
\includegraphics[width=0.75\textwidth]{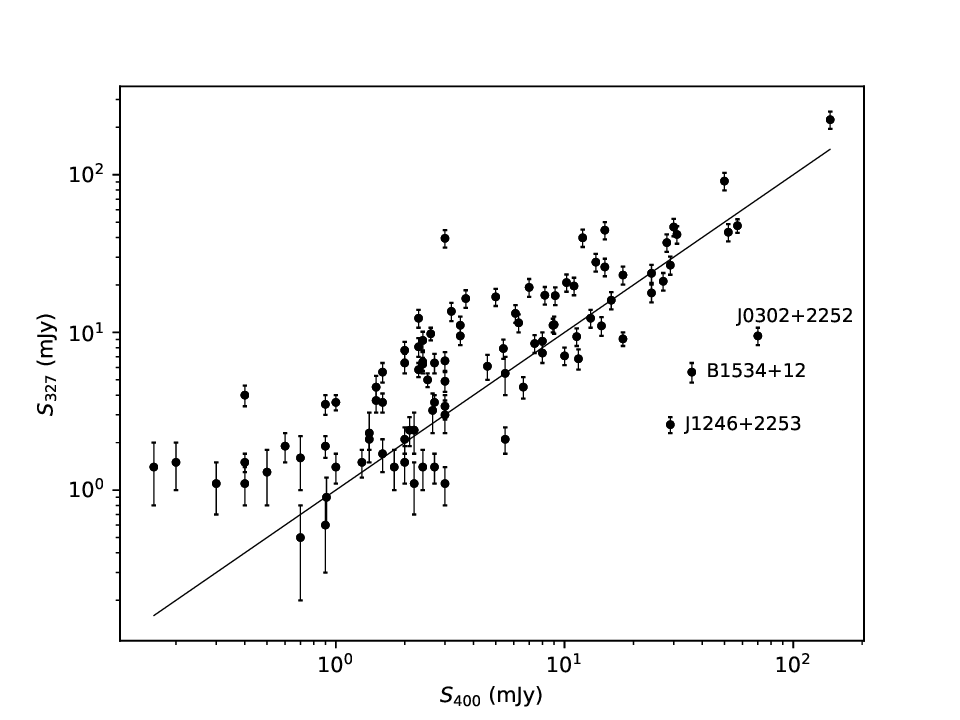}
\caption{Flux density for pulsar detections at 327~MHz calculated using Equation~\ref{eqn:s} vs. flux density at 400~MHz from the ATNF catalog. The solid line denotes $S_{\rm 327} = S_{\rm 400}$. The three labeled detections are outliers due to nulling (J0302$+$2252), scintillation (B1534$+$12), and an apparent outlier $S_{\rm 400}$ value (J1246$+$2253). We discuss these points in more detail in Section~\ref{sec:flux}.
  \label{fig_s400}}
\end{center} 
\end{figure}

Table~\ref{tab:fluxes} lists our flux density measurements. If a pulsar was detected more than once, $S$ was calculated and listed separately for each detection. {Several pulsars have $\delta$ reported only to the nearest degree in the ATNF catalog. In those cases the uncertainty in $S$ is too large for the flux density estimates to be meaningful and we exclude them from our results. Also excluded are several pulsars severely affected by RFI, because in those cases the RFI masking step described in Section~\ref{sec:rfi} significantly alters the sample statistics and makes the masked data unsuitable for calculating the flux density.

Figure~\ref{fig_s400} shows our flux density measurements at 327~MHz vs. measurements at 400~MHz from the ATNF catalog, if available. Three pulsars stand out as outliers with $S_{\rm 327} << S_{\rm 400}$: B1534$+$12, J0302$+$2252, and J1246$+$2253. J0302$+$2252 nulls on time scales of a few seconds to tens of seconds \citep{J0302}, and we also observe nulling in our detection of this pulsar. Because of our short integration time, nulling can explain the relatively low measured flux density at 327~MHz. B1534$+$12 is prone to scintillation and \cite{B1534} report a scintillation time scale of 11 minutes at 430~MHz. Scintillation is one possible explanation for the relatively low measured $S_{\rm 327}$ from our sole observation of this pulsar. {Another possible explanation is that, being a member of a relativistic double neutron star system, PSR~B1534+12 has been shown to go through changes in flux density and the pulse profile shape caused by geodetic precession \citep{Fonseca14}.}
To investigate further we use the {\tt pulsar\_spectra} code of \cite{Swainston22} which incorporates a large database of pulsar flux density measurements and finds a best fit by trying a number of spectral models of varying complexity. Our measurement of the flux density of B1534$+$12 at 327~MHz is 5.6(8)~mJy, in good agreement with the {\tt pulsar\_spectra} estimate for the same frequency, 7(1)~mJy. Similarly, for J1246$+$2253 \citep{Brinkman18} we find that our measurement of 2.6(3)~mJy is consistent with the {\tt pulsar\_spectra} estimate, 2.1(5)~mJy. The 400~MHz flux density measurements listed in the ATNF catalog for B1534$+$12 and J1246$+$2253 are significant outliers from the best fits obtained by {\tt pulsar\_spectra} for the spectra of the two pulsars based on measurements at multiple frequencies. }


\section{Non-detections}\label{sec:non}

Because some strips of sky were covered more than once or adjacent strips may overlap slightly in terms of the contour of FWHM beam coverage, there may be more than one data chunk covering a pulsar's position. If there are two or more data chunks covering the same pulsar, they may yield only detections or only non-detections or a mix of both. There are 174 known pulsars within the sky area covered by AO327 that were not detected. Another 12 pulsars whose positions were covered more than once have a mix of detections and non-detections in survey data. Here we examine the potential reasons for non-detections.

In a subset of the non-detection cases there is not enough information to reliably determine whether the pulsar should have been detected. For example, 57 of our non-detections have no flux density measurements listed in the ATNF catalog. Furthermore, 41 of our non-detections have position uncertainties larger than the AO327 beam radius and an unknown fraction of them may be in fact outside the telescope beam. Some non-detections are affected by both of these factors simultaneously.
 
Since the effective gain $\epsilon G$ depends on the pulsar's $\delta$ and track through the beam, we cannot compute one detection threshold $S_{\rm min}$ for the whole survey. Another {pulsar-dependent} factor affecting detectability is the observed pulse width $W_{\rm obs}$, which depends on the intrinsic pulse width and dispersion and scatter broadening. Due to these factors we calculated a separate detection threshold for each data chunk where a known pulsar passed through the beam. Then we found $S/S_{\rm min}$, where for detections $S$ was calculated according to Equation~\ref{eqn:s}. {For non-detections $S$ was estimated using the radio spectrum fitting code of \cite{Swainston22} in the cases where flux density measurements were available at several frequencies. For the remaining pulsars we estimated $S$ by fitting the power law $S \propto \nu^{\alpha_{\rm p}}$ to the flux densities of each pulsar obtained from the ATNF catalog for observing frequencies from 0.1 to 3~GHz.} For pulsars where there is only one ATNF flux density measurement we used the average spectral index calculated based on values listed in the ATNF catalog: $\alpha_{\rm p} = -1.78$ for pulsars with $P > 30$~ms and $\alpha_{\rm p} = -1.85$ for pulsars with $P < 30$~ms. The {pulsar-dependent} detection threshold is
\be
S_{\rm min} = \frac{\beta~\left( S/N\right)_{\rm min}~T_{\rm sys}}{\epsilon~G~\sqrt{N_p~\Delta\nu~T_{\rm obs}}} \sqrt{\frac{W_{\rm obs}}{P-W_{\rm obs}}}, \label{eqn:smin}
\ee
where $\left( S/N\right)_{\rm min} = 6$ is the candidate signal-to-noise threshold used in blind searches, and $W_{\rm obs}$ is the observed pulse width. For detected pulsars $W_{\rm obs} = W_{\rm 50}$ from Table~\ref{tab:pulsewidths}. 
At lower frequencies dispersion and scatter broadening cannot be ignored, especially in the case of MSPs. For non-detections, we calculated the observed pulse width based on the intrinsic pulse width $W_{\rm int}$ and broadening effects as
\be
W_{\rm obs} = \sqrt{W^2_{\rm int} + \Delta t^2_{\rm DM,1ch} + \Delta t^2_{\rm samp} + \tau^2_{\rm s}}
\ee
where $\Delta t_{\rm samp} = 81.92~\mu$s is the sampling time, $\Delta t_{\rm DM,1ch} = 8.3~\mu{\rm s}~DM~\Delta\nu_{\rm 1ch,MHz}/\nu^3_{\rm GHz} = 5.8~\mu{\rm s}~DM$. We assumed $W_{\rm int} = 0.1~P$
and obtained the scatter broadening time $\tau_s$ from the DM and the empirical relation in \cite{Bhat04}.

\begin{figure}
\begin{center}
\includegraphics[width=0.75\textwidth]{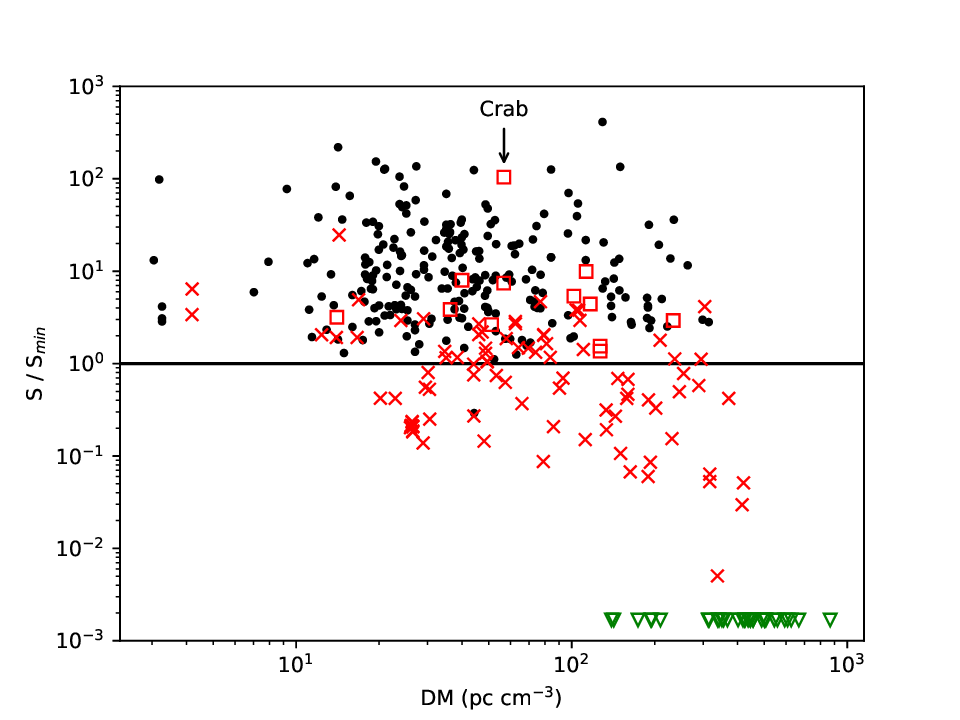}
\caption{Black dots show the ratio of flux density $S$ to detection threshold $S_{\rm min}$ vs DM for pulsar detections.
Red squares show the same for non-detections where $S$ was estimated using the pulsar spectrum fitting code of \cite{Swainston22}, and red crosses show the ratio for non-detections where $S$ was estimated by a power-law fit to flux density measurements from the ATNF catalog. Section~\ref{sec:non} details how $S$ and $S_{\rm min}$ were calculated for detections and non-detections. A solid line corresponds to the $6\sigma$ detection threshold commonly used in pulsar searches. Our weakest detection is below this threshold as known pulsars can sometimes be visually identified in diagnostic plots even if they would not pass the automated candidate sifting in blind searches, where a lower threshold would result in an overwhelming number of spurious candidates.
Green triangles indicate the DMs of non-detections for which $W_{\rm obs} > P$ in Equation~\ref{eqn:smin}. In those cases a threshold $S_{\rm min}$ cannot be calculated because their pulses are too broadened to be detected regardless of flux density.\label{fig_smin}}
\end{center} 
\end{figure}

Figure~\ref{fig_smin} shows $S/S_{\rm min}$ vs DM and it illustrates some of the causes for non-detections of known pulsars in our data. Most non-detections are at $DM > 50$~pc~cm$^{-3}$, indicating that dispersion and scatter broadening play a significant role. For a subset of non-detections $S_{\rm min}$ cannot be meaningfully calculated because $W_{\rm obs} > P$, i.e. their pulsed emission is not detectable as such regardless of flux density. These non-detections all have $DM > 100$~pc~cm$^{-3}$ and are close to or within the Galactic plane. AO327 did not target the Galactic plane but the survey's status as a filler project means that occasionally it covered strips of sky crossing the Galactic plane and populated with many highly dispersed known pulsars (Figure~\ref{fig_coverage}). Another set of non-detections are pulsars whose broadened pulse width does not exceed the rotation period but is nevertheless large enough that $S/S_{\rm min} < 1$. The increase in the number of non-detections (shown as red squares and red crosses) with increasing DM in the lower right of Figure~\ref{fig_smin} illustrates this cause for non-detection of known pulsars. 

Our non-detection of the Crab pulsar in data taken when it passed through the beam has a different cause. The Crab is among the brightest pulsars known and it is the sole non-detection with $S/S_{\rm min} > 100$ in Figure~\ref{fig_smin}. Inspection of the Crab data chunk revealed abnormal data statistics. The Crab was not in the beam when the backend was rebalanced at the start of the 1-hour observation containing its data chunk and saturated the backend's dynamic range when it came into the beam. The Crab nebula has an agular size of $6\arcmin \times 4\arcmin$, smaller than the $15\arcmin$ diameter of the Arecibo beam at 327~MHz. \cite{Bietenholz97} find that the nebula's flux density has a power-law dependence on observing frequency such that  $S_{\rm CN} \propto \nu^{-0.27}$. \cite{DeLooze19} report that the integrated flux density of the nebula is $\sim 834$~Jy at 1.4~GHz. Based on these values we obtain for the nebula an integrated flux density of $\sim 1235$~Jy at 327~MHz, the likely culprit for the backend saturation. The Crab pulsar \emph{was} detected both before it entered the main beam and after it exited as it did not cause saturation when it was detected in the sidelobes. We include these detections in the catalog but they are not included in Figure~\ref{fig_smin} as modelling sidelobes is beyond the scope of this paper. 


\section{Summary}\label{sec:conclusion}

The AO327 survey has detected 206 known pulsars listed in the ATNF catalog as periodic emitters. We have established a publicly accessible online repository of survey data and data products for these detections that can be used as a research and teaching resource for pulsar astronomy, signal processing and periodic signal detection algorithm development and testing, including artificial intelligence training. We hope to inspire researchers working on pulsar surveys with other observatories to similarly share detections of known pulsars and jointly establish a searchable online repository of pulsar data products. {We will explore options for making all raw data from the AO327 survey publicly available in the future as well.}

Data products from the surveys conducted with the Arecibo telescope over its half-century lifetime are an important scientific legacy and more recent data sets like AO327 are still a starting point for new research projects. The AO327 pulse profile catalog is a work in progress. We will continue to update the online incarnation of the catalog as more new pulsars are discovered and published by the AO327 team and others. 

\vspace{0.5cm}
J.S.D is supported by NSF award AST-2009335. M.A.M., E.F.L., and T.E.O. are supported by NSF award AST-2009425. M.A.M. is also supported by NSF Physics Frontiers Center award \#2020265. We thank Paul Ray for useful discussions about the intricacies of phase-aligning pulsar profiles across energy bands. We also thank Sebastien Guillot for providing the machine-readable ephemeris of PSR~J0751+1807 used in making Figure~\ref{fig_0751}. 

The Arecibo Observatory was a facility of the National Science Foundation operated under cooperative agreement by the University of Central Florida and in alliance with Universidad Ana G. Mendez, and Yang Enterprises, Inc. 

\begin{deluxetable}{lccrrll}
\label{tab:pulsewidths}
\tablecaption{Pulse widths $W$ and fractional duty cycles DC at 50\% and 10\% of the highest pulse peak, as determined from the single- or multi-Gaussian template fitted to the folded pulse profile of each pulsar. For pulsars with more than one detection the template was fitted to the folded pulse profile of the detection with the highest signal-to-noise. {The $P$ and DM columns contain values from the ATNF catalog except for the pulsars and respective parameters listed in Table~\ref{tab:dms}; in those cases the improved AO327 values are included here.} We list the first ten rows to illustrate the table's contents. The full table is provided in machine-readable form.}
\tablehead{ \colhead{Pulsar} & \colhead{P} & \colhead{DM} & \colhead{W$_{\rm 50}$} & \colhead{W$_{\rm 10}$} & \colhead{DC$_{\rm 50}$} & \colhead{DC$_{\rm 10}$}  \\
\colhead{} & \colhead{(s)} & \colhead{(pc~cm$^{-3}$)} & \colhead{(ms)} & \colhead{(ms)} & \colhead{} & \colhead{} }
\startdata
B0045+33 &  1.217094  &  39.92  & 18.4 &  33.9 &  0.015  & 0.028 \\
B0301+19 &  1.387584  &  15.66  & 48.1 &  84.0 &  0.035  & 0.061 \\
B0523+11 &  0.354438  &  79.42  & 14.7 &  22.3 &  0.042  & 0.063 \\
B0525+21 &  3.745539  &  50.87  & 64.0 & 214.0 &  0.017  & 0.057 \\
B0531+21 &  0.033392  &  56.77  &  2.0 &  7.3  &  0.061  & 0.218 \\
B0609+37 &  0.297982  &  27.15  &  6.3 &  18.8 &  0.021  & 0.063 \\
B0611+22 &  0.334960  &  96.91  & 11.0 &  19.8 &  0.033  & 0.059 \\
B0626+24 &  0.476623  &  84.18  &  8.4 &  22.3 &  0.018  & 0.047 \\
B0751+32 &  1.442349  &  39.99  & 30.3 &  71.8 &  0.021  & 0.050 \\ 
B0820+02 &  0.864873  &  23.73  & 22.4 &  40.1 &  0.026  & 0.046 \\
\enddata
\end{deluxetable}

\begin{deluxetable}{cclcccrr}
\label{tab:fluxes}
\tablecaption{A list of pulsar detections with MJD epoch and the raw data file containing the detection. The angular offset between the pulsar and beam center in declination $y$, gain correction factor $\epsilon$, $6\sigma$ detection threshold $S_{\rm min}$, uncalibrated flux density $S$, and flux density uncertainty $\sigma_{\rm S}$ are calculated separately for each detection since they depend on the pulsar's track through the beam. Some detections made outside the main beam are not shown in this table, although they appear in Table~\ref{tab:pulsewidths} (e.g. B0531+21 and B0751+32). We do not report flux density estimates for those cases as they are not covered by the representation of the main beam gain in Equation~\ref{eqn:epsilon}. Here we list the first ten rows to illustrate the table's contents. The full table is provided in machine-readable form.}
\tablehead{ \colhead{Pulsar} & \colhead{MJD} & \colhead{Raw data file basename} & \colhead{$y$} & \colhead{$\epsilon$} & \colhead{S$_{\rm min}$} & \colhead{S} & \colhead{$\sigma_{\rm S}$} \\
\colhead{} & \colhead{} & \colhead{} & \colhead{(\amin)} & \colhead{} & \colhead{(mJy)} & \colhead{(mJy)} & \colhead{(mJy)} }
\startdata
    B0045+33 & 58052 & 4bit\_puppi\_58052\_Strip225\_0482.0001  &  2.77  &  0.81  &  0.21  &  7.6  &  1.0 \\
    B0045+33 & 58101 & 4bit\_puppi\_58101\_Strip225\_1505.0001  &  2.85  &  0.80  &  0.21  &  4.9  &  0.7 \\
    B0301+19 & 58237 &  4bit\_puppi\_58237\_Strip88\_0562.0001  &  0.03  &  0.85  &  0.32  & 21.1  &  2.7 \\
    B0525+21 & 56787 &  4bit\_puppi\_56787\_Strip75\_1143.0001  &  3.43  &  0.50  &  1.47  & 47.5  &  4.7 \\
    B0609+37 & 58037 & 4bit\_puppi\_58037\_Strip238\_0282.0001  &  3.14  &  0.79  &  0.27  & 16.0  &  2.0 \\
    B0611+22 & 58931 &  4bit\_puppi\_58931\_Strip72\_0430.0001  &  4.32  &  0.35  &  1.04  & 26.7  &  3.5 \\
    B0626+24 & 57112 & 4bit\_puppi\_57112\_Strip954\_0532.0001  &  4.53  &  0.69  &  0.33  & 41.8  &  5.3 \\
    B0820+02 & 58977 & 4bit\_puppi\_58977\_Strip121\_0187.0001  &  6.99  &  0.46  &  0.44  & 46.6  &  5.9 \\
    B0919+06 & 58915 & 4bit\_puppi\_58915\_Strip100\_0142.0001  &  4.89  &  0.62  &  0.32  & 43.2  &  5.5 \\
    B1530+27 & 58186 &  4bit\_puppi\_58186\_Strip29\_0870.0001  &  4.39  &  0.69  &  0.34  & 12.3  &  1.6 \\
\enddata
\end{deluxetable}

\appendix

\section{Catalog Files Provided Online}\label{sec:catfiles}

The AO327 catalog is available online at {\tt http://ao327.nanograv.org}. At the top level in the file tree, there is a directory for each detected pulsar. One level down there is a sub-directory for each separate detection with a name following the format: {\bf 4bit\_puppi\_[MJD]\_Strip[Strip \#]\_[Raw File \#]\_D[HHMMSS+DDMM]}. The coordinates following 'D' are the $\alpha$ and $\delta$ of the middle of the data chunk extracted for that pulsar detection. Here we list the file types contained in each detection sub-directory:

{\bf *.fits :} The data chunk extracted for this pulsar detection, in 4-bit PSRFITS format.

{\bf *rfifind.mask :} RFI mask made by {\tt rfifind} in binary format. 

{\bf *rfifind.ps :} PostScript image of the RFI mask. 

{\bf *rfifind.inf :} Header information for the RFI mask in ASCII format. 

{\bf *rfifind.bytemask/.rfi/.stats :} Diagnostics for the RFI mask in a binary format compatible with PRESTO executables.

{\bf *[PSR name].dat :} Dedispersed time series in binary float format made with the ATNF DM or the better AO327 DM value from Table~\ref{tab:dms}, if the latter exists for the pulsar. This file is typically used with its corresponding *.inf file, listed next. 

{\bf *[PSR name].inf :} Header information of the dedispersed time series in ASCII format. This file is typically used in conjunction with its corresponding *.dat file above. 

{\bf *.pfd :} Binary file made by {\tt prepfold} containing  diagnostics in a format compatible with PRESTO executables. 

{\bf *.pfd.ps/.png :} A PostScript and a PNG image of the diagnostics contained in the corresponding *.pfd file. 

{\bf *.pfd.bestprof :} Folded pulse profile in ASCII format; pulse phase relative to the start of the observation. 

{\bf *.pfd.bestprof.zerophase :} Folded pulse profile in ASCII format rotated to absolute pulse phase according to the procedure described in Section~\ref{sec:phaseref}. This pulse profile can be aligned with other folded pulse profiles by using the absolute phase reference provided in the corresponding *phaseref.txt file (described below).

{\bf [PSR name].gauss :} Analytical single- or multi-peaked Gaussian pulse template in ASCII format made by {\tt pygaussfit.py}.

{\bf [PSR name].gauss.template.zerophase :} Binned pulse profile template in ASCII format made by evaluating the corresponding analytical [PSR name].gauss template and then rotating it with {\tt zerophase.py} according to the procedure described in Section~\ref{sec:phaseref}.

{\bf *phaseref.txt :} ASCII text file listing the {\tt get\_TOAs.py} command used to calculate the absolute phase reference, and the phase reference itself in the form of a TOA as well as TZRSITE/TZRFRQ/TZRMJD parameters that can be used with an existing pulsar ephemeris to align folded pulse profiles from other data with the AO327 folded pulse profile from the corresponding *pfd.bestprof.zerophase file. 

\software{PRESTO\footnote{\tt https://www.cv.nrao.edu/\~{}sransom/presto} \citep{Ransom02}, {\tt pulsar\_spectra}\footnote{\tt https://pulsar-spectra.readthedocs.io} \citep{Swainston22}}

\end{document}